\documentclass[twocolumn]{aastex63}
\usepackage{amsmath}
\usepackage{amssymb}
\usepackage{graphicx,float}
\usepackage[utf8]{inputenc}
\usepackage{booktabs}
\usepackage{array,multirow}
\usepackage{gensymb}

\begin{document}

\title{\large Contribution of Secondary Neutrinos from Line-of-sight Cosmic Ray Interactions to the IceCube Diffuse Astrophysical Flux }

\author{Alina Kochocki}
\affiliation{\textit{Department of Physics and Astronomy, UCLA, Los Angeles, CA 90095, USA}}
\author{Volodymyr Takhistov}
\affiliation{\textit{Department of Physics and Astronomy, UCLA, Los Angeles, CA 90095, USA}}
\affiliation{\textit{Kavli Institute for the Physics and Mathematics of the Universe (WPI), UTIAS \\The University of Tokyo, Kashiwa, Chiba 277-8583, Japan}}
\author{Alexander Kusenko}
\affiliation{\textit{Department of Physics and Astronomy, UCLA, Los Angeles, CA 90095, USA}}
\affiliation{\textit{Kavli Institute for the Physics and Mathematics of the Universe (WPI), UTIAS \\The University of Tokyo, Kashiwa, Chiba 277-8583, Japan}}
\author{Nathan Whitehorn}
\affiliation{\textit{Department of Physics and Astronomy, Michigan State University, E. Lansing, MI 48824, USA}}
\affiliation{\textit{Department of Physics and Astronomy, UCLA, Los Angeles, CA 90095, USA}}

\received{December 9th, 2020}
\accepted{April 13th, 2021}
\begin{abstract}
In ten years of observations, the IceCube neutrino observatory has revealed a neutrino sky in tension with previous expectations for neutrino point source emissions. Astrophysical objects associated with hadronic processes might act as production sites for neutrinos, observed as point sources at Earth. Instead, a nearly isotropic flux of astrophysical neutrinos is observed up to PeV energies, prompting a reassessment of the assumed transport and production physics. This work applies a new physical explanation for neutrino production from populations of active galactic nuclei (AGN) and starburst galaxies to three years of public IceCube point source data. Specifically, cosmic rays (CRs) produced at such sources might interact with extragalactic background light and gas along the line of sight, generating a secondary neutrino flux. This model is tested alongside a number of typical flux weighting schemes, in all cases the all-sky flux contribution being constrained to percent levels of the reported IceCube diffuse astrophysical flux.
\end{abstract}

\section{INTRODUCTION}

In the last decade, IceCube has established the existence of high energy, astrophysical neutrinos \citep{Aartsen_2013a}. While this spectrum is observed as a significant excess over the dominating atmospheric neutrino background, no source class or mechanism has been found to substantially contribute to this flux. In 2018, a single point source was identified. This blazar AGN, TXS 0506+056, was associated with one high energy neutrino event ($\sim$ 300 TeV) coincident with a gamma-ray flare. Analysis of historical events in the blazar's direction revealed a neutrino flaring state from September 2014 through March 2015 (a negligible contribution to the overall diffuse intensity). Due to its relative distance, this blazar was not favored as a likely neutrino source prior to detection. \citep{147}. 

AGN, starburst galaxies and other dynamic environments capable of proton acceleration, have all been proposed as potential neutrino sources through pion decay \citep{M_sz_ros_2017}. Bolstered by this theoretical expectation, and the result of TXS 0506+056, multiple studies have attempted to account for this observed diffuse flux assuming direct emissions from well-motivated source populations, only to constrain these contributions to fractional levels ($\sim 20\%$)  \citep{Aartsen_2020,Aartsen_2019,Aartsen_2017a,Hooper_2019,smith2020revisiting}. 

This observed isotropy and lack of sufficient directional correlation with populations of nearby, likely primary neutrino sources, could be the first evidence for neutrino production along the line of sight, an emission model offering increased preference for the contributions of distant sources. Blazars and other AGN are expected to accelerate protons to ultrahigh energies. Protons with energies above $10^{17}$~eV can interact with extragalactic background light (EBL) photons and produce neutrinos in the reaction $p + \gamma_{b} \rightarrow n + \pi^{+},$ followed by pion decay. There is growing evidence that such interactions are responsible for the high-energy spectra of the most distant blazars~\citep{Essey_2009,Essey_2010,Essey_2011,Essey:2011wv,Razzaque:2011jc,Murase_2012,Essey:2013kma,Aharonian:2012fu,Prosekin:2012ne,Inoue:2013vpa,DeFranco:2017wdr,Archer:2018ehn,Gueta:2019mxx,Das:2019gtu,Toomey:2020jjz}.  Neutrinos produced in the same interactions can account for the $\sim$PeV neutrinos observed by IceCube~\citep{Kalashev_2013}. Furthermore, interactions of high-energy protons with gas, especially in clusters intervening along the line of sight, can produce neutrinos for lower proton energies in $pp$ collisions followed by the decay of mesons. The signature of secondary production is deviation from the usual inverse square scaling of flux with distance. Specifically, this production would enhance the relative contributions of the furthest, numerous sources, increasing the expected level of isotropy.

We develop a framework for using IceCube data to identify and model the neutrinos produced in secondary interactions along the line of sight to distant sources of cosmic rays. We evaluate the fluxes of secondary neutrinos and compare the results with the predictions of a primary production hypothesis. 

\section{SECONDARY NEUTRINOS FROM LINE-OF-SIGHT INTERACTIONS}

There is little doubt that AGN jets can accelerate protons, and they are widely considered the likely sources of ultrahigh-energy cosmic rays with energies up to $10^{11}$~GeV. Theoretical models support the acceleration of protons to at least $10^8$~GeV~\citep{Sironi:2010rb,Sironi:2013ri}, and some exceptional conditions allow for an even higher energy. These protons can cross cosmological distances with little loss of energy and can generate high-energy gamma rays in their interactions with cosmic background photons and protons (gas) along the line of sight. As long as the extragalactic magnetic fields are below $\sim$10 femtogauss, in agreement with observations~\citep{Neronov:1900zz,Ando_2010,Dermer:2010mm,Essey_2010nd,Chen:2014rsa,Chen:2018mjd}, the trajectories of protons in excess of 100 TeV are not deflected significantly outside the stronger magnetic fields of galaxy clusters \citep{Lee_1995}. While the local galactic magnetic fields cause significant deflections, most of the neutrinos produced along the line of sight point back to their origin, allowing association with a distant source. In fact, gamma-ray observations are consistent with magnetic fields as low as $10^{-18}$~G~\citep{Dermer:2010mm,Essey_2010nd}, and the lower bound depends on the unknown duty cycle of blazars~\citep{Dermer:2010mm}. 

The relevant targets for neutrino production include universal photon backgrounds, such as EBL and CMB (cosmic microwave background), as well as hydrogen, which is present in intergalactic space, but much more abundantly in clusters of galaxies. For a distant source, the line of sight has {\cal O}(1) probability to pass without crossing any galaxies or clusters~\citep{Berezinsky:2002vt,Aharonian:2012fu}, and {\cal O}(1) probability to be interrupted by a cluster of galaxies~\citep{Berezinsky:2002vt,Aharonian:2012fu}. For unobstructed lines of sight, EBL is the main target. This target is optically thin (only a small fraction of protons interact with EBL), suggesting an optical depth proportional to distance. The flux of secondary neutrino scales as the proton flux, $F_{\textrm{proton}}$, multiplied by the distance, $D$: 
\begin{equation}
F_{\textrm{secondary}, \nu} \propto F_{\textrm{proton}} \times D \propto \dfrac{1}{D}. 
\end{equation}
This is different from $F_{\textrm{primary}, \nu}\propto 1/D^2$, which we will use to test the origin of the neutrinos statistically. 

The scaling is more complicated for the lines of sight that cross at least one cluster. The main target is now likely to be the cluster, not the entire length of the line of sight. Additionally, the proton trajectory is likely to be deflected significantly by the strong magnetic fields that can be found in clusters. In this case, the scaling of secondary neutrino flux with distance is complicated, but still approaches $1/D$.  

Line-of-sight interactions attributable to a combination of both background photon fields, and gas distributed throughout clusters, filaments and voids, provide a general mechanism for the production of secondary neutrinos spanning the relevant energy range of IceCube. As there is large uncertainty in the intervening matter density on a source-by-source basis, exact modeling of the expected neutrino spectrum for a given population is not considered. Instead, this work tests the hypothesis that secondary interactions play a dominant role in accounting for the observed diffuse spectrum. Neutrino events of energies below 100 TeV are  mainly due to \textit{pp} collisions, carrying off only a fraction of the initial cosmic ray energy, while interactions with background photon fields yield higher energy neutrinos. We also note, it is possible that a combination of both primary and secondary neutrino production contributes to IceCube's diffuse astrophysical spectrum. In this study, the possible distinct contributions from both neutrino production mechanisms are explored in modeling the signal expectations from the populations of blazar and non-blazar AGN, as well as starburst galaxies.

\section{MODELING THE EMISSIONS OF NEUTRINO POINT SOURCE POPULATIONS OBSERVED BY ICECUBE}

IceCube consists of 86 vertical cables, or strings, each holding sixty optical modules (hemispherical photomultiplier tubes encased in a glass housing, pressurized to less than 1 atm). The strings are frozen into the Antarctic ice, instrumenting a cubic kilometer of target material. Neutrinos which interact within this detector volume or the surrounding medium produce relativistic charged particles: an electron, muon or tau corresponding to the flavor of the neutrino. The propagation of each lepton through the ice is characterized by its signature emission pattern of Cerenkov radiation, detected at the optical modules \citep{Aartsen_1998}.

This study utilizes three years of public IceCube data (2010, 2011 and 2012)\footnote{https://icecube.wisc.edu/science/data/PS-3years}. The set consists of approximately 350,000 candidate neutrino events. Specifically, this sample reports track-like events, generally produced in muon neutrino charged current interactions \citep{Aartsen_2017a}. The pointed nature of these tracks is ideal for angular reconstruction and directional astronomy, having an angular error of $\sim$0.5$ \degree$. This study uses the reported right ascension (R.A.), declination ($\delta$), and angular error ($\sigma$), of each reconstructed event. Effective area as a function of zenith angle, energy, and year (varying string configurations), is also utilized as reported by the IceCube collaboration.

Taking into account systematic uncertainties involved with IceCube angular reconstruction, the distribution of neutrino events from a point source is observed as a 2D Gaussian (on a sphere), at Earth. The signal PDF of reconstructed event direction $\vec{x} _{i}(\delta_{i}, \textrm{R.A.}_{i})$, angular error, $\sigma _{i}$, and source location, $\vec{x} _{s}$ is modeled as \citep{Aartsen_2017a,aartsen_1999,Hooper_2019,smith2020revisiting}:
\begin{equation}
    S(\delta_{i}, \textrm{R.A.}_{i}) = \dfrac{1}{2 \pi \sigma _{i}^{2}}   \textrm{exp} \Bigg( - \dfrac{1}{2}  \bigg( \dfrac{ | \vec{x} _{i} - \vec{x} _{s} |  }{ \sigma _{i} }   \bigg)^{2} \Bigg). 
\end{equation}

The signal likelihood for source location, $\vec{x} _{s}$, is the product of $N$ individual neutrino event PDFs within some surrounding region under consideration:
\begin{equation}
    \mathcal{L} (n_{s}) = {\displaystyle \prod_{i=1}^{N} } \:   \bigg[ \dfrac{n_{s}}{N} S(\delta_{i}, \textrm{R.A.}_{i}) + (1 - \dfrac{n_{s}}{N})B(\textrm{sin}(\delta_{i})) \bigg].
\end{equation}
We optimize this value with respect to the number of signal events, $n_{s}$. $B$ represents a constant neutrino background PDF, taken as $1/\Omega$ for the case of a uniform sky ($\Omega$, the solid angle spanned by the included region). In this analysis, a band of $\pm 6 \degree$ around an individual source or test location is considered. 

Notably, the event rate of the data used for this work shows a high level of zenith dependence. To account for this variation, the original $\pm 6 \degree$ band is split into twelve sub-bands, the kth sub-band of solid angle $\Omega_{k}$, containing N$_{k}$ events. The generalized background PDF for this region becomes:
\begin{equation}
\textit{B}(\textrm{sin}(\delta_{i})) = \dfrac{N_{k}}{N \times \Omega_{k} }, 
\end{equation} 
 which reduces to $1/\Omega$ for a true uniform background.

The maximizing number of signal events, $n_{s,\textrm{max} }$, and optimized likelihood product with respect to background:
\begin{equation}
\lambda = 2 \cdot (\textrm{ln}(\mathcal{L}) \{n_{s} = n_{s,\textrm{max} }    \} - \textrm{ln}(\mathcal{L})\{ n_{s} = 0  \} ), 
\end{equation}
may both serve as test statistics, gauging the presence of a source at some test location. This study utilizes the minimum variance estimate provided by $\lambda$ in analysis. An unbinned search steps through the observed space in small increments, testing all regions for non-Gaussian fluctuations in this value.

\begin{figure*}[t]
\centering
\includegraphics[width=\linewidth]{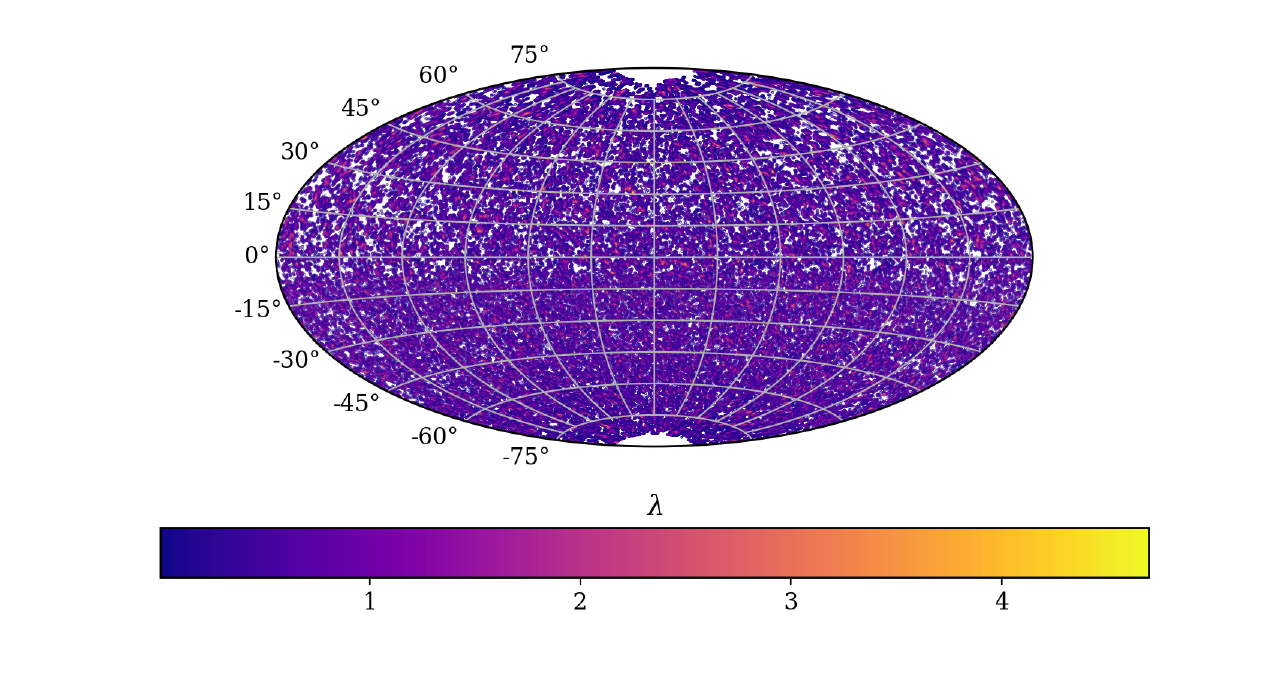}
\caption{Point Source Grid Scan Sky Map. Pictured above (Aitoff projection), are the results of an all-sky search for point sources in three years of IceCube reconstructed track (muon), events. The square root of the test statistic, $\lambda$, is shown per test location. The value of the square-root test statistic appears randomly distributed with location, generally peaked at a value of zero (dark purple). No obvious position can be associated with a point source. }
\end{figure*}

Beyond examining individual test points or known source locations, a joint likelihood, $\mathcal{L}_{pop}$, for a population of astrophysical sources may be considered \citep{aartsen_1999}:
\begin{equation}
\begin{aligned}
\textrm{ln}(\mathcal{L}_{pop} \{ n_{s} \}) = {\displaystyle \sum_{i=1}^{N}}  \textrm{ln} \Bigg(\dfrac{n_{s}}{N} \cdot  S(\delta_{i}, R.A._{i}, \sigma_{i}) \\
+ \bigg(1 -  \dfrac{n_{s}}{N} \bigg) \cdot B(\textrm{sin}(\delta_{i}) ) \Bigg). 
\end{aligned}
\end{equation}
 Here, $n_{s}$ represents the cumulative signal event contribution from the population. All neutrino events contained within the data set are considered for their association with the set of $N_{src}$ sources. The signal PDF becomes a weighted product of individual source signal PDFs:
\begin{equation}
\textrm{S}(\delta_{i}, \textrm{R.A.}_{i}, \sigma_{i}) = \dfrac{\sum_{j=1}^{N_{src}} w_{j} \cdot S_{j}(\delta_{i}, \textrm{R.A.}_{i} , \sigma_{i} )}{ \sum_{j=1}^{N_{src}} w_{j}}. 
\end{equation}
The signal PDF, $S_{j}$, for the jth source location, is determined from equation (2) for the ith neutrino event. The individual weight, $w_{j}$, is chosen to represent some expectation for a source's contribution to the total observed neutrino flux. In general, this weight will be the product of a detector specific component informed by the effective area, and a second component dependent on the emission model proposed for the flux expectation. 

The weight accounting for the reported detector sensitivity is determined by integrating effective area with a chosen energy distribution. A power law of index either $\alpha =$ 2.0 or 2.5 is chosen. The value of 2.0 is motivated by the prediction of Fermi acceleration at the source. An index of 2.5 closely matches the best fit power law results of IceCube \citep{Aartsen_2015, Aartsen_2025}. This effective area weight is calculated as a function of source zenith angle, $\theta$:
\begin{equation}
w_{ \textrm{Aeff}}(\theta, \alpha ) = \int_{E_{min}}^{E_{max}} \textrm{Aeff}(E,\theta) \times E^{-\alpha} dE.
\end{equation}
Here, neutrino energy, $E$, runs from 100 GeV to 10$^{9}$ GeV. 

Next, one of four hypotheses for the dominant emission model is considered:

\begin{itemize}
    \item \textbf{Line-of-Sight (Secondary) Production Weighting:} The flux from a member of a certain astrophysical source class scales as $1/D$, where $D$ is the luminosity distance. To account for redshift dependence,     $\Lambda$CDM cosmology is assumed with cosmological parameters: $H_{0} = 67$ km s$^{-1}$ Mpc$^{-1}$, $\Omega_{\Lambda} = 0.68$, and $\Omega_{M} = 0.32$ \citep{Planck_2020}. The distance is determined below as:
    \begin{equation}
    D = \dfrac{c(1+z)}{H_{o}} \int_{0}^{z} dz' [ \Omega_{M} (1 + z' )^{3} + \Omega_{\Lambda}  ]^{-1/2}.
    \end{equation}
    In this case, the weight assigned to each source would be:
    \begin{equation}
        w_{\textrm{model, Secondary}} = \dfrac{1}{D}.
    \end{equation}
    \item \textbf{Flat Weighting:} All sources are modeled as producing the same flux of neutrinos at Earth, $w_{\textrm{model, flat}} = 1$. No additional information is used to physically motivate the weight. This model provides a conservative approach to testing the possibility of neutrino emission associated with a given population. 
    \item \textbf{Gamma-Ray Weighting:} In this case, the reported integral gamma-ray flux of a source is used as its weight, $\phi$. This tests the hypothesis that primary gamma rays, neutrinos (and CRs), are produced in some proportion, $w_{\textrm{model, $\gamma -$Ray}} = \phi $.
    \item \textbf{Geometric Weighting:} This weighting follows the assumption that a flux of primary neutrinos falls off with the inverse square of distance, $w_{\textrm{model, Geometric}} = 1/D^{2}$.
\end{itemize}

The final weight for a $j$th source and weighting scheme is given as:
\begin{equation}
    w_{j} = w_{\textrm{Aeff}} \times w_{\textrm{model}}.
\end{equation}

\begin{figure}[!thb]
\centering
\includegraphics[width=\linewidth]{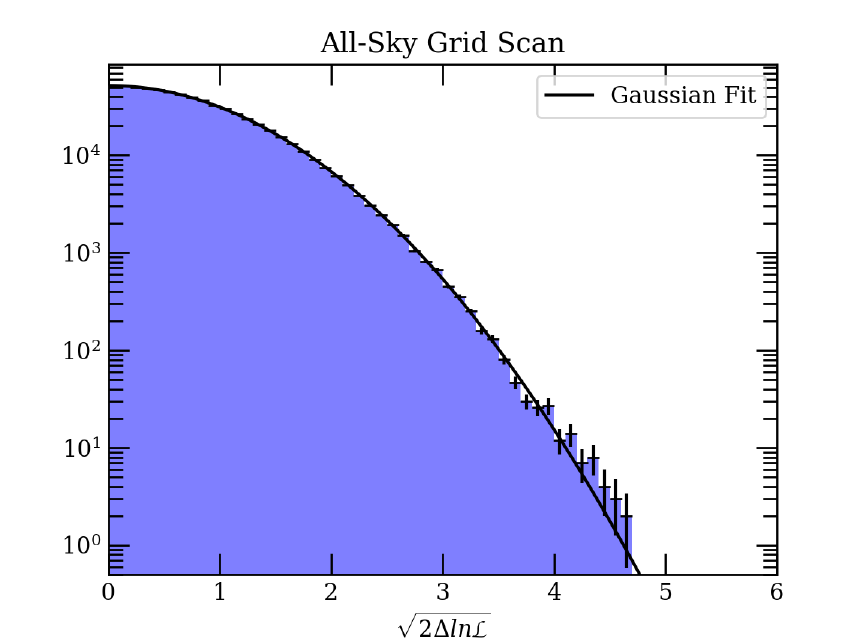}
\caption{Point Source Grid Scan Likelihood Distribution. The distribution of the square-root of the test statistic, $\lambda$, is displayed in blue for test locations of the all-sky grid scan. The Gaussian shape is expected from Wilks' Theorem.  }
\end{figure}

Finally, for a total number of neutrino events associated with a given population, $n_{s}$, the differential muon neutrino flux can be determined from the observation time, T, effective area and chosen energy distribution. Assuming each flavor is seen in equal proportion at Earth, the all-flavor diffuse flux is found by scaling the muon neutrino flux by three:
\begin{equation}
\begin{aligned}
    \Phi_{\nu} = 3 \dfrac{n_{s}}{T} \times \bigg( \int d\Omega  \int_{E_{min}}^{E_{max}}  \textrm{Aeff}(E,\theta) \\ \times E^{-\alpha} dE \bigg)^{-1}.
\end{aligned}
\end{equation}

This function of $n_{s}$ is utilized in later sections to explore the total flux contribution of a given population and weighting with respect to the reported IceCube diffuse spectrum. 

\section{ALL-SKY POINT SOURCE GRID SCAN}
As a preliminary measurement, an all-sky grid scan is performed to identify any prominent individual point sources. Right ascension and declination are stepped through in $0.2 \degree$ increments. Each location is tested for the presence of a point source signal through maximizing equation (3). Test locations of declination within six degrees of zenith or nadir are excluded due to low event rates in these regions. 
The sky map of the square root of the test statistic, $\lambda$ is shown in Figure 1. Only test locations associated with a physical, positive number of signal events are shown. The corresponding distribution is pictured in Figure 2 with Poisson uncertainties, fit by a Gaussian of mean zero. The distribution appears normal, without any significant deviation indicative of a signal. These results agree with those of other works \citep{Aartsen_2020,Aartsen_2019,Aartsen_2017a,Hooper_2019,smith2020revisiting}. 

In the event the contribution of individual sources is insignificant, the overall flux from a given source class may provide sufficient evidence for a detection. In the next sections, the contributions of different subsets of active galactic nuclei from the Fermi 4LAC catalogue, and a set of starburst galaxies are considered.  

\section{BLAZAR \& NON-BLAZAR AGN}

The Fermi 4LAC catalogue draws upon eight years of observations of active galactic nuclei as reported in the recent Fermi 4FGL source catalogue, in addition to other pertinent multi-instrument data culled from an expansive literature search \citep{latcollaboration2020fourth}. The 4LAC catalogue contains 2863 objects of high Galactic latitudes ($|b| > 10 \degree$). In each case, the location in right ascension and declination, object class (distinguishing between blazar and non-blazar AGN), integral photon flux from 1-100 GeV, and a variability index (gauging gamma-ray variability), are provided in addition to other information. 

In general, the number of sources used for different weighting schemes within a population will vary depending on the amount of available information per source. In this case, approximately half of the 4LAC AGN are reported with a measured redshift: 999 well-determined redshifts taken from the BZCAT catalogue and other optical surveys, and 656 with possible photometric contamination or other noise contributions. 

\begin{figure}[!t]
\centering
\includegraphics[width=\linewidth]{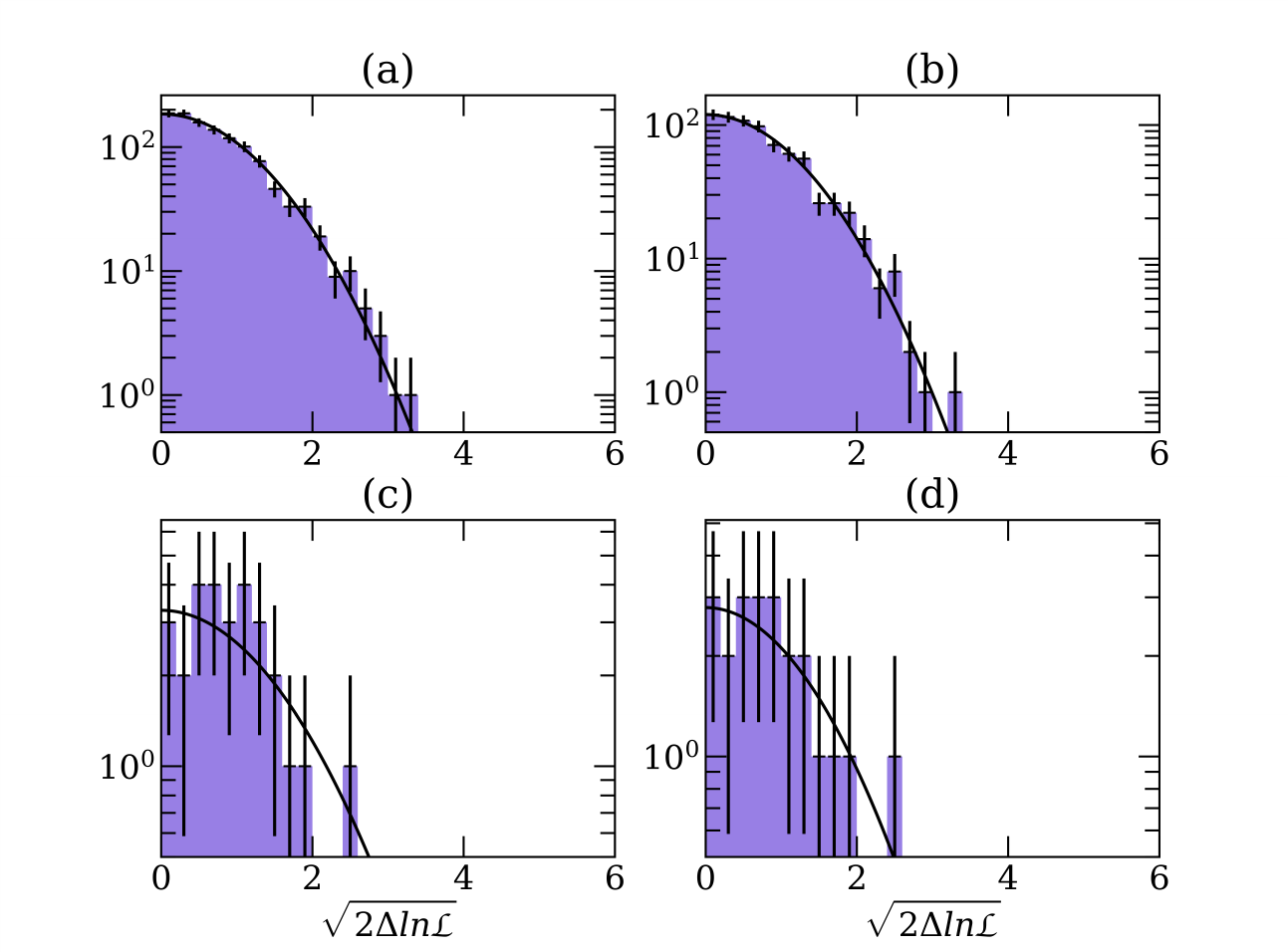}
\caption{AGN Likelihood Distributions. Shown above are distributions of the square root of the likelihood test statistic at locations specific to the four AGN subpopulations of this work. Each is shown with a Gaussian fit and Poisson counting uncertainties. The figures (a) and (b) represent the locations of all 2796 blazar AGN and 1674 nonvariable blazar AGN, respectively. Similarly, figures (c) and (d) represent the results of non-blazar AGN and nonvariable, non-blazar AGN. Each distribution appears normal in shape, suggesting these studies have measured statistical fluctuations from the null result (background).  }
\end{figure}

As there is some possibility of mis-associating AGN with local stellar objects, sources with measured redshifts corresponding to distances of $< 1$ Mpc have been excluded from subpopulations considered for EBL and geometric weightings. In particular, two supposed blazar AGN have been removed (4FGL J0654.0-4152 and 4FGL J0719.7-4012). Sources of declination, $87 \degree < |\delta|$, are also excluded due to a decreased amount of solid angle at the poles.

The set of blazar AGN is considered first, a commonly proposed source for UHECR \citep{Murase_2012}. Three 4LAC AGN classes are included within this set: Flat Spectrum Radio Quasars, BL Lacertae objects, and blazar candidates of unknown type. In total, 2796 objects are considered for their contribution to the diffuse flux in flat and gamma-ray weighting schemes, and 1531 sources are used for weighting models reliant on distance. The distribution of the square-root likelihood test statistic determined at all 4LAC blazar locations is presented in the top right plot of Figure 3. 

\begin{figure*}[!thb]
\gridline{\fig{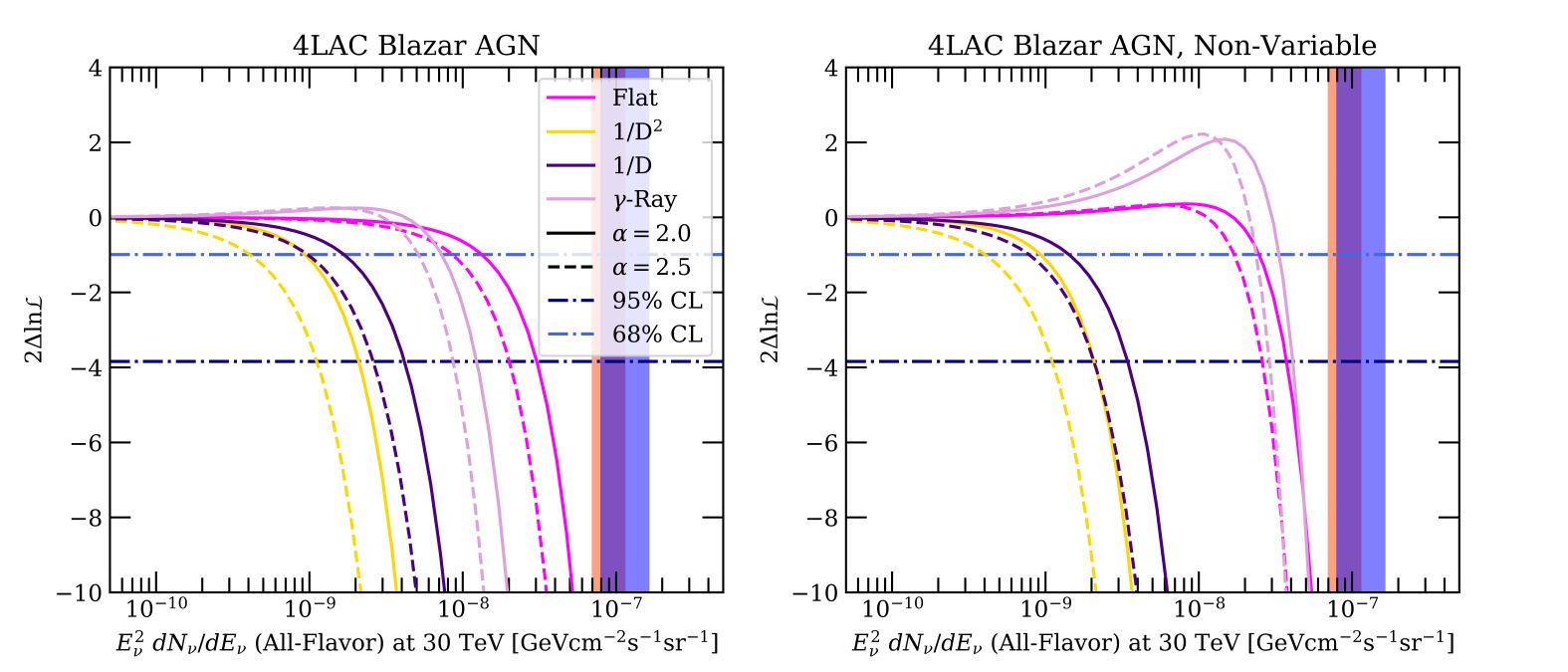}{\textwidth}{(a) Pictured above are the likelihood test statistics for populations of 4LAC blazar AGN (left), and nonvariable, blazar AGN (right), as a function of the associated 30 TeV, all-flavor flux, equation (12). Four different emission models are tested with two neutrino energy distribution indexes. The horizontal blue dash-dotted lines indicate 68\% and 95\% confidence levels. Vertical blue and red transparent regions indicate the 68\% confidence bands for the measured 30 TeV all-flavor diffuse flux (reported by IceCube in 2015 and 2020, respectively \citep{Aartsen_2015, Aartsen_2025}). As a general feature of this work, it should be noted that a smaller subpopulation does not necessarily provide a lower constraint on the related flux contribution. A larger population will provide a more constraining model, this effect potentially competing with the presence of a real signal. }}
\gridline{\fig{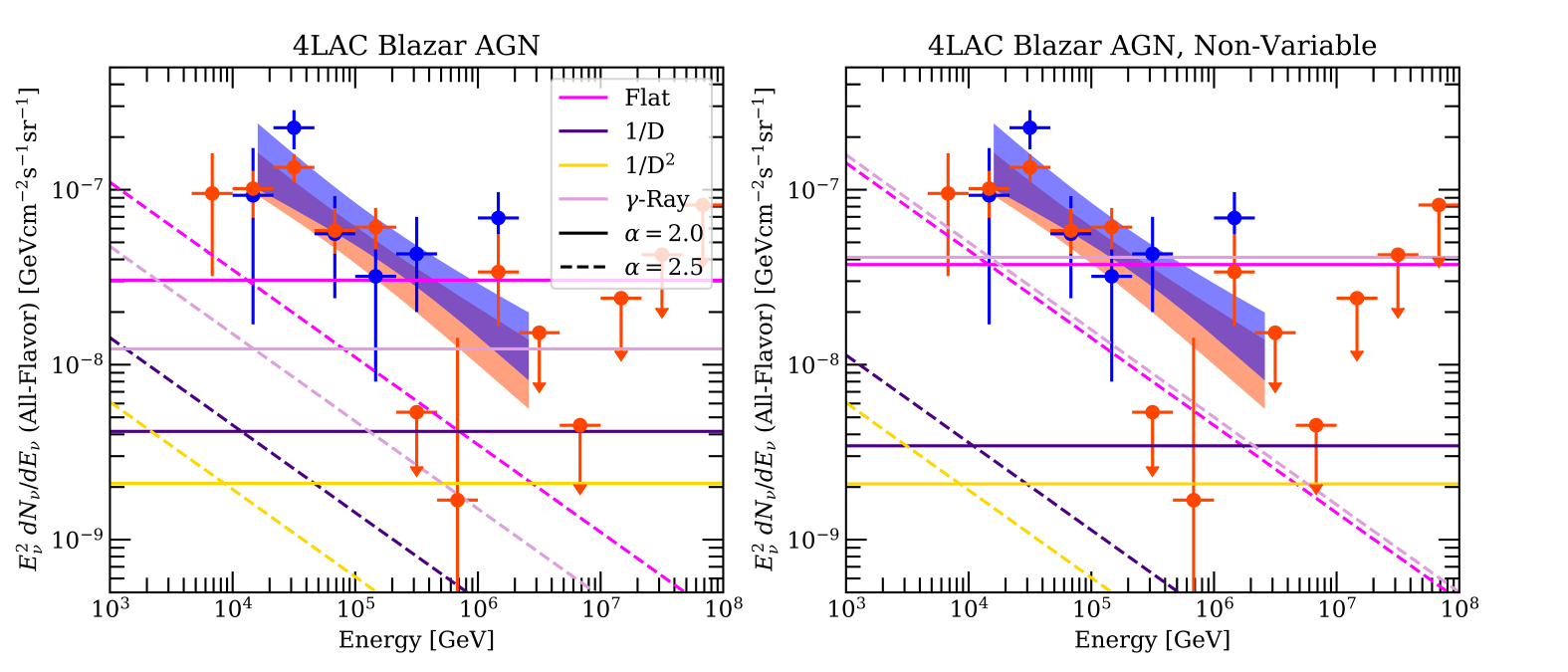}{\textwidth}{(b) The plots above show power law spectra corresponding to the 95\% confidence level 30 TeV flux of each weighting and index. This value is determined from the results of the likelihood space scan shown in Figure 4, (a). These spectra represent the neutrino emissions associated with each weighting. Blue and red transparent bands correspond to power law spectra describing the observed all-flavor diffuse flux, published by IceCube \citep{Aartsen_2015, Aartsen_2025}. Blue and red markers indicate binned spectral measurements from either study.}}
\caption{Blazar AGN All-Flavor Flux Likelihood Space. In the above figures, the diffuse, all-flavor flux at 30 TeV for eight weighting schemes is mapped as a function of likelihood. No significant detections are made. Considering the 95\% upper limits shown in (b), neither population can account for the observed diffuse flux. }
\end{figure*}

\begin{figure*}
\gridline{\fig{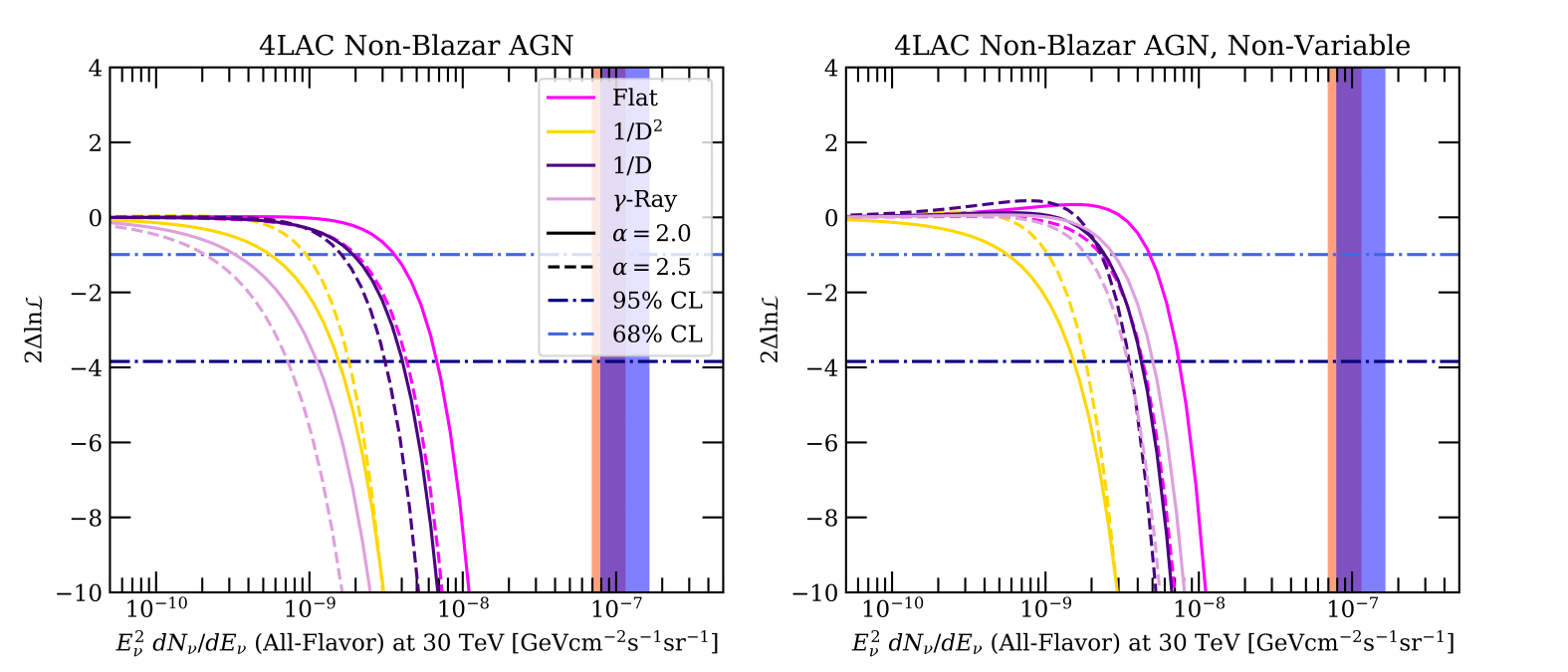}{\textwidth}{(a) Shown above are the likelihood space scans for the population of 4LAC non-blazar AGN (left), and nonvariable, non-blazar AGN (right). Eight different weights are tested (four emission models with two indexes for the assumed neutrino energy distribution). Transparent blue and red bands show the the 68\% confidence regions for the 30 TeV all-flavor diffuse flux (IceCube, 2015 and 2020, respectively).}}
\gridline{\fig{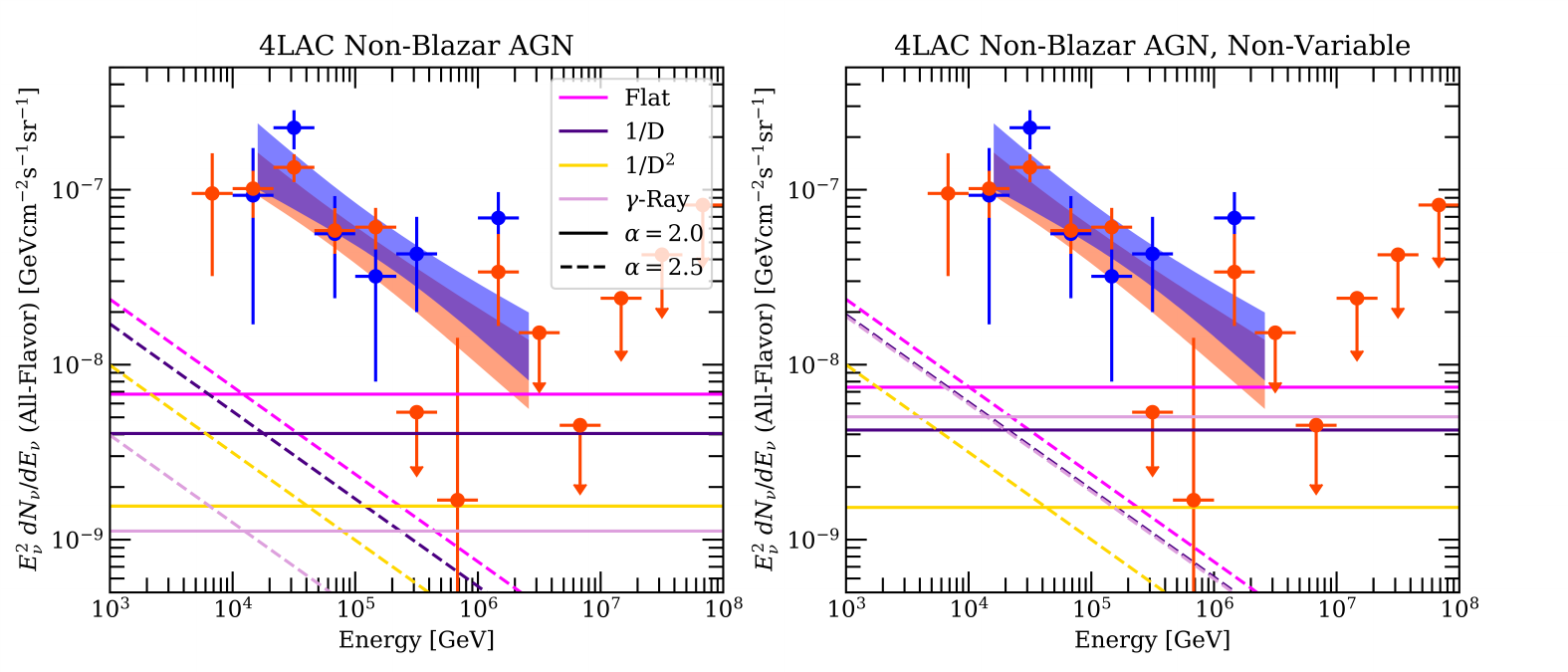}{\textwidth}{(b) Pictured above are power law energy distributions representative of the 95\% confidence level 30 TeV flux as determined in Figure 5, (a), for each weighting. These spectra reflect the neutrino signals associated with each emission model and energy distribution ($\alpha = 2.0, 2.5$). Blue and red power law bands correspond to 68\% confidence levels determined by IceCube (2015 and 2020). The colored markers (blue and red), represent binned flux measurements of either study.}}
\caption{Non-Blazar AGN All-Flavor Flux Likelihood Space. The 30 TeV diffuse, all-flavor fluxes accounted for by the weighting schemes of this study are shown above as a function of likelihood. There are no significant detections. Referencing (b), no population is found to account for the observed diffuse flux at the 95\% confidence level.}
\end{figure*}

Next, the total signal contribution from the combined population of blazar AGN is considered for eight possible weighting schemes (four emission models each paired with the effective area weight for one of two indexes). The likelihood of equation (6) is determined as a function of $n_{s}$, then converted to an all-flavor neutrino flux following equation (12). Curves of this likelihood space corresponding to fractional levels of the reported IceCube diffuse flux are pictured in the upper left plot Figure 4. Here, a positive test statistic corresponds to a greater likelihood for the data to be described by a positive signal component than to be described by background alone (null result, $n_{s} = 0$). A negative test statistic suggests the data is better described by pure background than with the addition of a given signal component. Following Wilks' Theorem, the 68$\%$ and 95$\%$ confidence limits are expressed in terms of the likelihood test statistic. The neutrino signal potentially associated with each emission model and energy distribution is also shown for comparison with IceCube's observed all-flavor diffuse flux in the bottom left of Figure 4.

Similar to \citep{smith2020revisiting}, the subpopulation of blazar AGN with low levels of gamma-ray variability is also considered. This requirement is met by using only those blazars of reported variability index less than 18.48. Ultimately, 1674 sources are used for the flat and gamma-ray weighting scenarios, and 767 sources are considered for weights dependent on distance. The resulting distribution of square-root test statistics for all 1674 independent locations is provided the top right of Figure 3. 

The likelihood space for this population of nonvariable blazar AGN is mapped for each of eight weighting schemes. The results of this scan and the neutrino emission spectra related to each weight are compared with the reported diffuse flux of IceCube in Figure 4.

A small percentage (64 sources), of non-blazar AGN are also included within the 4LAC catalogue. Following \citep{smith2020revisiting}, the two lobes of Centauras A are decidedly treated as a single source, and the additional sources Centaurus B and 3C 411 are included with redshifts. No gamma-ray measurement is assumed for either source. The final population consists of 65 sources to be used for the flat weighting, 63 with reported gamma-ray photon fluxes, and 60 sources sufficient for geometric and EBL studies. 

Similarly, the population of nonvariable, non-blazar AGN may be studied. 47 such sources are considered, of which 45 have gamma-ray photon measurements, and 43 have measured redshifts.

The square-root test statistic likelihood distributions for both non-blazar AGN and nonvariable, non-blazar AGN are pictured in the lower half of Figure 3. The result of each likelihood space scan and the neutrino emissions corresponding to each weight for both non-blazar AGN and nonvariable, non-blazar AGN are shown in Figure 5. 

\begin{figure*}
\gridline{\fig{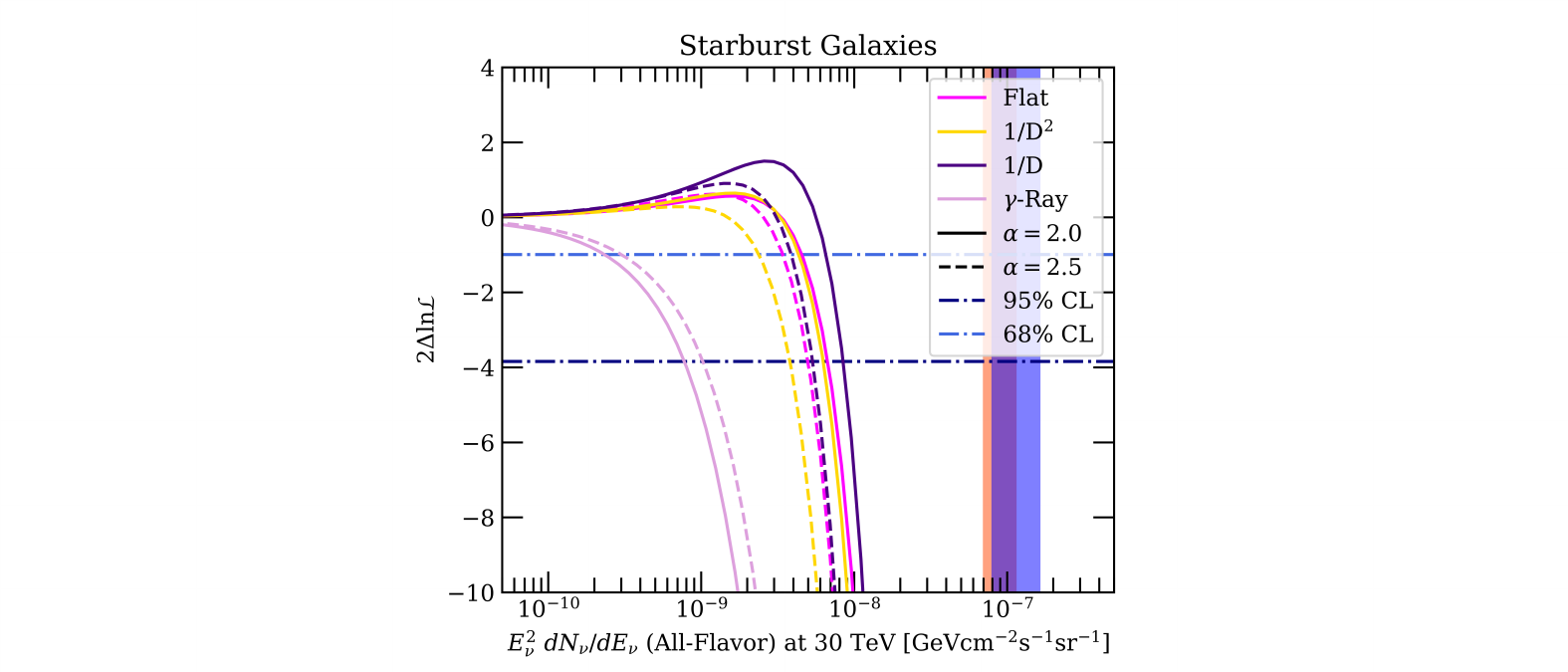} {\textwidth}{(a) Pictured above are the likelihood space scans for a population of starburst galaxies. As in Figure 4 and Figure 5, the emission models presented in this work are used to determine the overall flux contribution from this source population. The 68\% confidence level 30 TeV all-flavor diffuse flux is represented by blue and red horizontal bands, as reported by IceCube in 2015 and 2020 \citep{Aartsen_2015, Aartsen_2025}.}}
\gridline{\fig{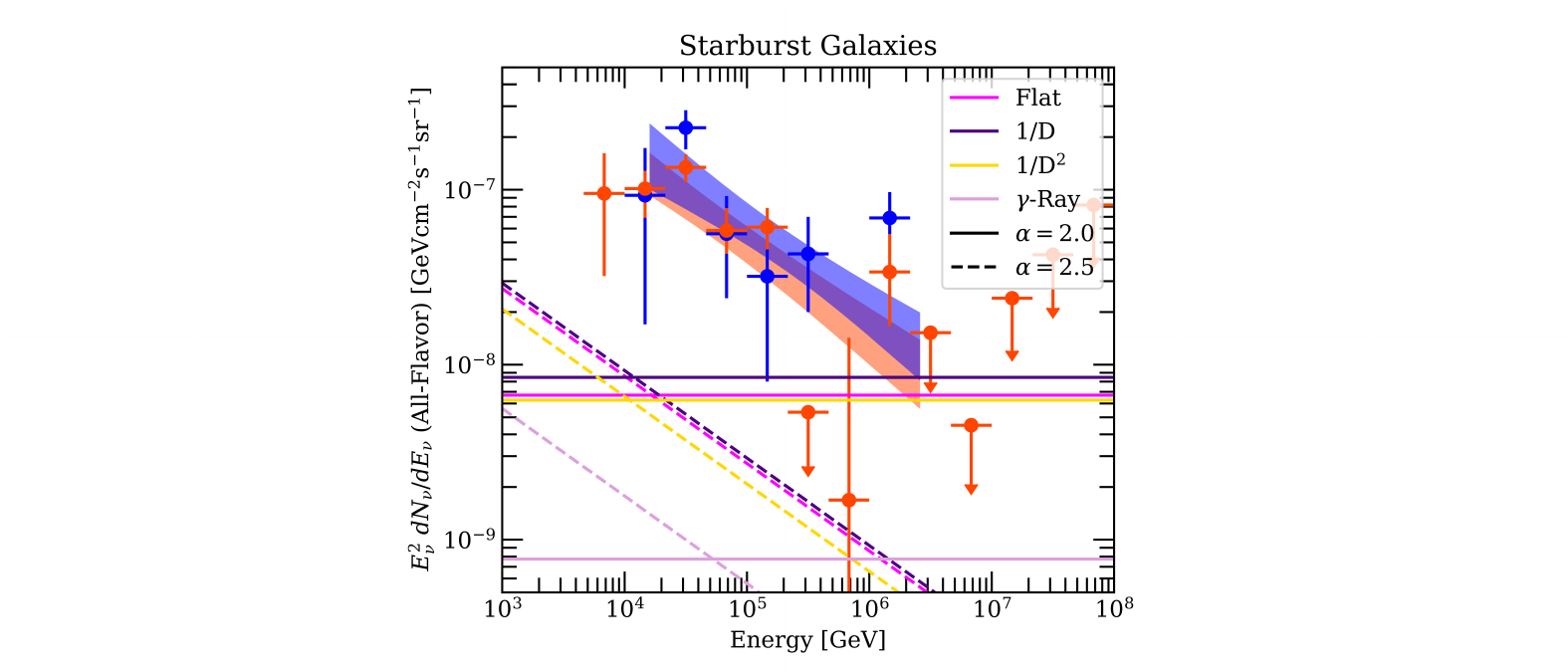}{\textwidth}{(b) Shown above are the power law neutrino energy distributions (of index 2.0  and 2.5), with normalizations determined by the 95\% 30 TeV flux upper limits of Figure 6, (a). These spectra represent the neutrino emissions corresponding to each tested weight. The blue and red markers represent binned spectral flux measurements determined by IceCube, and the corresponding power law fits at the 68\% confidence level (2015 and 2020).}}
\caption{Starburst Galaxies All-Flavor Flux Likelihood Space. The figures above show the all-flavor flux likelihood space for eight weighting schemes, and the corresponding power law energy distributions determined by the 95\% flux upper limit. No significant detections are found. At most, starburst galaxies could account for only a small fraction of the IceCube all-flavor diffuse flux. }
\end{figure*}

\section{STARBURST GALAXIES}

The possible contribution of starburst galaxies to the observed diffuse neutrino flux is also considered. Starburst galaxies, regions associated with a high rate of star formation, are another source class commonly associated with the production of UHECR \citep{PhysRevD.97.063010,Zaw_2009}. A table of 45 local, radio-bright starburst galaxies is referenced (1.4 GHz flux density greater than 0.3 Jy) \citep{Lunardini_2019}. For consistency with the earlier AGN analyses of this work, only the 43 sources at distances greater than 1 Mpc are considered, all of which have consistently reported infrared flux densities (60 $\mu $m), and distances. In this study, an infrared luminosity, $L_{IR}$, is determined for each source of flux density, $S_{60 \mu \textrm{m}} $, and distance, $D$. This result is converted to a gamma-ray flux using the following empirical relation:
\begin{equation}
    \textrm{log}_{10} \bigg( \dfrac{L_{\gamma}}{( \textrm{erg s}^{-1} )} \bigg) = \alpha \textrm{log}_{10} \bigg( \dfrac{L_{IR}}{( 10^{10}L_{\odot})}
 \bigg) + \beta.
\end{equation}

Here, $L_{\gamma}$ is gamma-ray source luminosity and $L_{\odot}$ is the solar luminosity. Values of $\alpha = 1.18 \pm 0.15$ and $\beta = 38.49 \pm 0.24$ have been referenced from the literature \citep{PhysRevD.96.083001}. The resulting likelihood scan for this population and associated neutrino emission spectra are presented in Figure 6.

\renewcommand{\arraystretch}{1.0}
\begin{table*}[!thb]
\begin{center}
\begin{tabular}{|c|c|c|c|c|}\hline
  Source Class & Index & Emission Model & $E^{2} dN_{\nu}/dE_{\nu}$  at 30 TeV 68\% UL & $E^{2} dN_{\nu}/dE_{\nu}$ at 30 TeV 95$\%$ UL\\ \hline
\multirow{8}{*}{Blazar AGN} & \multirow{4}{*}{2.0} & Flat &  $<$ 13.3\phm{123} & $<$ 30.4\phm{123}  \\ 
    &  & Secondary  & 1.7  & 4.2  \\
    &  & Geometric & 0.9 & 2.1  \\
    &  & Gamma-Ray & 7.3 & 12.3\phm{1}  \\ \cline{2-5}
    & \multirow{4}{*}{2.5} & Flat & 8.5 & 20.2\phm{1}  \\
    &  & Secondary & 1.0  & 2.6  \\
    &  & Geometric &  0.4   & 1.1  \\
    &  & Gamma-Ray  & 5.1  & 8.7   \\\cline{2-5} \hline
\multirow{8}{*}{Blazar AGN, Nonvariable} & \multirow{4}{*}{2.0} & Flat & 24.7\phm{1}  & 37.4\phm{1}  \\ 
    &  & Secondary   & 1.4   & 3.4  \\
    &  & Geometric & 0.9  & 2.1  \\
    &  & Gamma-Ray & 33.4\phm{1}  & 41.1\phm{1} \\ \cline{2-5}
    & \multirow{4}{*}{2.5} & Flat  &  17.0\phm{1}  & 26.0\phm{1}  \\ 
    &  & Secondary  & 0.8  & 2.1 \\
    &  & Geometric &  0.4   & 1.1  \\
    &  & Gamma-Ray  & 23.7\phm{1} & 28.9\phm{1}   \\ \hline

\multirow{8}{*}{Non-Blazar AGN} & \multirow{4}{*}{2.0} & Flat & 3.6  & 6.8 \\ 
    &  & Secondary   & 1.9  & 4.1  \\
    &  & Geometric & 0.6  & 1.6 \\
    &  & Gamma-Ray & 0.3  & 1.1 \\ \cline{2-5}
    & \multirow{4}{*}{2.5} & Flat &  2.0  & 4.3  \\ 
    &  & Secondary & 1.6  & 3.1 \\
    &  & Geometric &  1.0 & 1.8 \\
    &  & Gamma-Ray  &  0.2  & 0.7   \\ \hline
    
\multirow{8}{*}{Non-Blazar AGN, Nonvariable} & \multirow{4}{*}{2.0} & Flat & 4.8  & 7.4  \\ 
    &  & Secondary     & 2.4  & 4.2 \\
    &  & Geometric  & 0.6  & 1.5  \\
    &  & Gamma-Ray & 2.8  & 5.0  \\ \cline{2-5}
    & \multirow{4}{*}{2.5} & Flat & 2.3  & 4.3 \\ 
    &  & Secondary & 2.3  & 3.5 \\
    &  & Geometric &  1.0  & 1.8  \\
    &  & Gamma-Ray & 1.9  & 3.5  \\ \hline

\multirow{8}{*}{Starburst Galaxies} & \multirow{4}{*}{2.0} & Flat & 4.5  & 6.7   \\
    &  & Secondary     & 6.5  & 8.5 \\
    &  & Geometric   & 4.3  & 6.3  \\
    &  & Gamma-Ray  & 0.2  & 0.8  \\ \cline{2-5}
    & \multirow{4}{*}{2.5} & Flat & 3.4  & 5.0  \\ 
    &  & Secondary & 3.9 & 5.3 \\
    &  & Geometric  & 2.4 & 3.8   \\
    &  & Gamma-Ray & 0.3  & 1.0 \\ \hline

\end{tabular} 
\caption{Diffuse Flux Contribution by Source Population - Upper Limits. In this table, the 30 TeV all-flavor diffuse flux 68\% and 95\% upper limits are given in units of $10^{-9}$ GeVcm$^{-2}$s$^{-1}$sr$^{-1}$ for each source population and weighting scheme. Here, a weighting scheme is determined by both an emission model, and the index assumed for the neutrino energy distribution. As reference, the 30 TeV all-flavor diffuse flux from the 2020 and 2015 IceCube analyses are 122$_{+25}^{-21} \times 10^{-9}$ GeVcm$^{-2}$s$^{-1}$sr$^{-1}$ and 94$_{+43}^{-43} \times 10^{-9}$ GeVcm$^{-2}$s$^{-1}$sr$^{-1}$ at the 68\% confidence level, respectively \citep{Aartsen_2025, Aartsen_2015}. No tested population or weighting scheme is found to fully account for the observed diffuse flux.}
\end{center}
\end{table*}

\section{RESULTS \& DISCUSSION}

Considering first the all-sky grid scan method, no significant point sources are identified. These results are consistent with other analyses, both of this point source data set \citep{Hooper_2019,smith2020revisiting}, and those performed on different or larger event samples by the IceCube collaboration \citep{Aartsen_2020,Aartsen_2019,Aartsen_2017a}. It is also notable that a distribution of the square root of the likelihood test statistic follows a Gaussian, consistent with Wilks' Theorem. 

As in the case of the all-sky study, each population-specific square-root test statistic distribution appears normal. This is true for blazar AGN as well as non-blazars and star-forming galaxies. There is no location motivated by association with an astrophysical source found with a significant neutrino signal in this work. 

Regardless of the contributions identified from individual sources, a well-motivated emission model might have the potential to properly emphasize the flux of certain sources, leading to a statistically significant signal from an entire population. Considering the maximum test statistic of each likelihood curve, there are no detections for any source class or weighting scheme within this study. The 68\% and 95\% flux (all-flavor) upper limits for each population and weighting scheme have been reported in Table 1. 

In the cases of non-blazar AGN, nonvariable non-blazar AGN and starburst galaxies, the 95$\%$ all-flavor flux upper limit is constrained to within 10$\%$ of IceCube's diffuse astrophysical flux, independent of emission model. The populations of blazar AGN and nonvariable blazar AGN cannot be ruled out as significant contributors, possibly associated with 30-40$\%$ of the diffuse flux at the 95$\%$ confidence level (flat and gamma-ray weighting schemes).

In all AGN populations of this work, the flat weighting 95\% and 68\% confidence levels regularly exceed those of the line-of-sight production weighting, which in turn exceed those of the geometric weighting. In the case of a perfectly isotropic event distribution (null result), a mean of $n_{s} = 0$, would be expected at any proposed source position. The geometric model would drive down the expected contributions of distant objects, only probing the flux of a small number of local sources. This model would provide the strongest constraints. The line-of-sight production weighting would stand one degree closer to the flat weighting scheme, physically assuming an equivalent contribution from each location (poorest constraints). 

The population of starburst galaxies considered in this work displays a deviation in the ordering of emission model upper limits predicted for the case of the null result. Specifically, the line-of-sight production model is found to have the least constraining upper limits at the 95\% and 68\% confidence levels, for energy distributions of indexes 2.0 and 2.5. While the peak test statistics of these two weighting schemes are clearly positive, the associated flux is not detected at a significant level. 

In this work, we have taken a phenomenological approach to explain the diffuse power law emissions observed by IceCube. Specifically, neutrino signals produced via the emission models of this work are assumed to mimic such an energy distribution. The case of a dominant neutrino contribution from secondary production is then tested by its characteristic flux weighting as a function of source distance: $1/D$. Considering the large range of suggested magnetic field strengths, 10$^{-18}$ to 10$^{-14}$ Gauss \citep{Neronov:1900zz,Ando_2010,Dermer:2010mm,Essey_2010,Chen:2014rsa,Chen:2018mjd}, and the considerable uncertainty in EBL and gas densities along a specific source's line of sight, secondary production is a feasible explanation even for the lowest energy astrophysical neutrinos. Our framework then allows us to consider the general family of models which might produce IceCube's observed diffuse spectrum through this production mechanism, without depending on assumptions of the energy spectrum or details or the line-of-sight-interaction mechanism; studies involving predictions of the energy spectrum would provide complementary constraints.

\section{CONCLUSION}

While this study reports no significant detections, it is possible the weighting schemes used in this and similar works greatly underestimate the physical complexity of emissions from a given source class. Specifically, the secondary production mechanism may require more sensitive consideration for proper modeling. Neutrino production is enhanced when a primary CR passes through a region of high gas and background light density in transit to Earth. Incorporating knowledge of intermediate matter density along the line of sight may provide the proper choice of weighting scheme. An ideal test would consider the density and location of galaxy clusters on the sky; these regions would be properly modeled by secondary neutrinos from an interaction with EBL or gas, while empty portions of the sky might yield direct neutrinos described physically by a combination of $1/D$, $1/D^2$, and gamma-ray models. In particular, such a correlation could be performed with thermal Sunyaev Zeldovich effect maps (y-maps). These maps describe the level of distortion of cosmic microwave background photons from inverse Compton scattering on the high energy electrons of galaxy clusters \citep{2016}.   

Proper modeling should also utilize neutrino event energy (not provided with this public set of data). This would allow for further differentiation between emission models. Direct neutrinos from astrophysical sources would correspond to lower energies, while events of energies beyond hundreds of TeV might be best described through secondary neutrino production \citep{Kalashev_2013}.

It is also notable that an improved model would have the ability to further probe the effect of the intergalactic magnetic fields (IGMF) on cosmic ray propagation. This improvement, accompanied by a larger IceCube data set and improved measurement of the EBL spectrum, would provide a novel opportunity to constrain the IGMF strength. Similarly, new constraints on the magnetic field strengths of primary neutrino sources may also be possible \citep{Irene}.

Increased consideration for the nature of specific source populations may also contribute to understanding of the diffuse flux. Specifically, allowing for variability in neutrino production to account for changes in source cosmic ray luminosity and column density could provide a more realistic model. This increased tolerance has been considered in previous work, demonstrating a greater primary neutrino flux contribution from starburst galaxies when variability in source luminosity is allowed \citep{ambrosone2020starburst}. It may also be important to consider the effect of unresolved astrophysical sources on the diffuse spectrum. The impact of such gamma-ray dim sources on the blazar luminosity function has been explored in \citep{Yuan:2019ucv}. In general, properly accounting for the complete luminosity function of blazars, or any class of astrophysical source, could enhance the associated neutrino flux \citep{Murase:2016gly}.  

Ultimately, additional investigation is necessary to better understand the interplay between different modes of neutrino production. While a more refined weighting model may better establish the presence of secondary neutrinos, we also leave open the possibility for an untested process or source class to significantly contribute to the IceCube diffuse spectrum. As starburst galaxies and AGN are arguably two of the most popular point source candidates suggested to play a role in cosmic ray and neutrino production, the dominant generator of the diffuse flux remains obscured.

\acknowledgments
We would like to thank Warren Essey, Andrew Ludwig, Hans Niederhausen, and Daniel Smith for conversations beneficial to this work. This research was generously funded in part by the UCLA College Honors Programs. A.K. and V.T. were supported by the U.S. Department of Energy (DOE) Grant No. DE-SC0009937 and by the World Premier International Research Center Initiative (WPI), MEXT, Japan. A.K. was also supported by Japan Society for the Promotion of Science (JSPS) KAKENHI Grant No. JP20H05853. We also acknowledge use of the Hoffman2 Shared Cluster computational and storage services provided by UCLA's Institute for Digital Research \& Education.

\nocite{*}
\bibliographystyle{aasjournal}
\bibliography{neutrino}
 
\listofchanges
\end{document}